\documentclass[10pt,conference]{IEEEtran}

\usepackage{cite}
\usepackage{amsmath,amssymb,amsfonts}
\usepackage{graphicx}
\usepackage{textcomp}
\usepackage{xcolor}
\usepackage{booktabs}
\usepackage{multirow}
\usepackage{url}
\usepackage{xspace}
\usepackage{algorithm}
\usepackage{algpseudocode}

\newcommand{\eg}{{\em e.g.},\xspace}

\newcommand{\tool}{\textsc{WaitLens}\xspace}

\begin{document}

\title{\tool: Temporal VLM-Based Adaptive Waiting for Industrial GUI Testing Agents}

\author{
\IEEEauthorblockN{
Shika Ikeura\IEEEauthorrefmark{1},
Xuan Wang\IEEEauthorrefmark{1},
Ziyi Kuang\IEEEauthorrefmark{1},
Jianping Xu\IEEEauthorrefmark{1},
Chao Kang\IEEEauthorrefmark{2},\\
Yongxiang Hu\IEEEauthorrefmark{1},
Xiaoyu Wang\IEEEauthorrefmark{2},
Juxing Yuan\IEEEauthorrefmark{2}, and
Yangfan Zhou\IEEEauthorrefmark{1}}
\IEEEauthorblockA{\IEEEauthorrefmark{1}College of Computer Science and Artificial Intelligence, Fudan University, Shanghai, China}
\IEEEauthorblockA{\IEEEauthorrefmark{2}Meituan, Beijing, China}
\IEEEauthorblockA{\{pcgchi25, xuanwang25, zykuang25, jpxu25\}@m.fudan.edu.cn, kangchao@meituan.com,\\
huyx@fudan.edu.cn, \{wangxiaoyu17, yuanjuxing\}@meituan.com, zyf@fudan.edu.cn}
}

\maketitle
\raggedbottom

\begin{abstract}
Improving the efficiency of GUI testing agents is important for industrial deployment.
Beyond LLM inference, fixed waiting between consecutive interactions introduces substantial overhead.
Our industrial study reveals that fixed waiting accounts for 42.3\% of the time within the sampled interaction intervals, comparable to the 54.8\% spent on model calls.
Moreover, 62.9\% of the waiting time is spent after the GUI has already finished loading.
Although prior studies have proposed adaptive waiting strategies, their reliance on app-specific data or platform-specific information limits their applicability across diverse industrial environments.
To this end, we propose \tool, which fine-tunes a lightweight VLM on GUI loading trajectories annotated with waiting end times.
We introduce temporal resampling to construct diverse training examples, enabling the model to learn varying GUI loading behaviors from temporal observations.
Experiments show that \tool achieves 86.08\% accuracy in deciding whether to keep waiting, compared with 54.13\% for the backbone VLM.
Our evaluation using recordings from Meituan's industrial testing pipeline shows that \tool reduces mean extra waiting after the interface is ready by 94.98\% compared with a 20~s fixed wait.
\end{abstract}


\section{Introduction}

\begin{figure*}[t]
    \centering
    \includegraphics[width=0.8\textwidth]{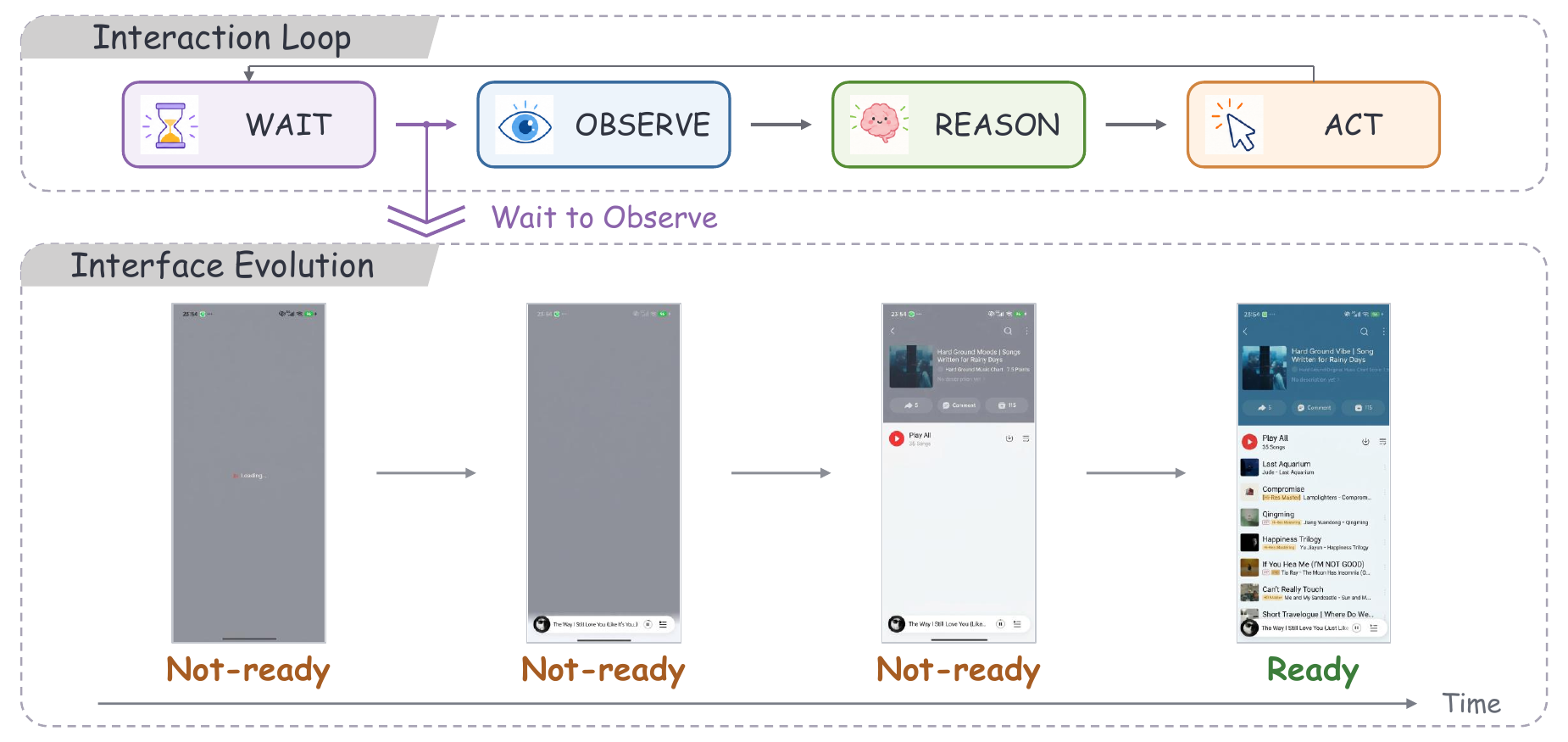}
    \caption{GUI testing agent interaction loop and interface evolution during post-action waiting. The upper panel shows the interaction loop, and the lower panel illustrates the interface progressing from intermediate states to Ready.}
    \label{fig:intro_interaction_loop}
\end{figure*}

Recently, large language model (LLM) based GUI agents have emerged as a new testing method, which reduces reliance on manually maintained test scripts~\cite{liu2024gptdroid,yoon2024droidagent,ran2024guardian,hu2024auitestagent,chen2025memodroid,kong2025prophetagent,feng2025agentforuser,wang2025testweaver}.
Given natural language testing tasks, these agents repeatedly observe GUI screenshots, reason about subsequent actions, and interact with apps to complete testing tasks.
As illustrated in Figure~\ref{fig:intro_interaction_loop}, each action is followed by a waiting period before the agent captures the next screenshot, allowing the interface to respond.

Determining an appropriate waiting time, however, is challenging.
If the wait is too short, agents may mistake an intermediate loading state for the result of the preceding action, leading to incorrect decisions~\cite{wu2025backtrackagent,wu2025guireflection}.
To accommodate slow responses, existing GUI agents often adopt fixed waiting times after each action~\cite{zhang2025appagent,wang2024mobileagent,wang2024mobileagentv2}.
However, a long fixed wait also keeps agents idle when the interface becomes ready earlier.
Our motivation study in Section~\ref{ssec:motivation_study} shows that 62.9\% of the waiting time in our industrial GUI testing agent is spent idling after the GUI has already finished loading.
This substantial overhead motivates adaptive waiting policies in GUI testing agents.

Although prior studies have explored adaptive waiting for GUI testing~\cite{feng2023adat,feng2023wereplay,zhang2024robustwaiting,li2026turbodroid}, these methods face limitations when applied to GUI agents across diverse industrial environments.
First, learning-based methods rely on app-specific training data and may struggle with unseen GUIs.
For example, AdaT~\cite{feng2023adat} uses a convolutional neural network (CNN) to classify GUI rendering states.
To adapt AdaT to WeChat, WeReplay~\cite{feng2023wereplay} collects new training data from the app and fine-tunes the model.
Second, analysis-based methods rely on internal runtime or platform-specific information that screenshot-based agents cannot directly access.
For example, Zhang et al.~\cite{zhang2024robustwaiting} monitor DOM changes, while TurboDroid~\cite{li2026turbodroid} uses method dependency analysis to determine waiting conditions.
Therefore, GUI agents need an adaptive waiting approach that relies only on screenshots and can generalize across different GUIs.

Vision-language models (VLMs) have demonstrated strong capabilities in understanding diverse GUIs~\cite{baechler2024screenai,you2024ferretui,li2025ferretui2,cheng2024seeclick}, making them a promising alternative.
However, directly applying existing VLMs to adaptive waiting is impractical for two reasons.
First, unlike static GUI tasks, judging whether an interface has finished loading requires observing how the screen changes over time~\cite{chen2025guiworld,lin2024videogui}.
Simply feeding all historical screenshots into a VLM increases input size and inference latency as loading continues.
This may offset the time saved by adaptive waiting.
Second, general-purpose VLMs are not specifically trained to distinguish loading states from completed interfaces.
As a result, they may struggle to accurately predict whether the GUI is ready.

To address these challenges, we propose \tool, an VLM-based approach for adaptive waiting in GUI testing agents.
Instead of waiting for a fixed duration, \tool periodically checks the screen after each action to decide when testing should proceed.
To capture interface changes within a bounded input size, \tool uses a sliding window of screenshots as input to model.
During VLM fine-tuning, we use temporal resampling to augment the training data and simulate different GUI loading speeds.
Together, these designs enable \tool to predict GUI readiness across diverse loading conditions with low inference overhead.

We conduct comprehensive experiments to evaluate the effectiveness of \tool.
Our in-house experiments show that \tool achieves 86.08\% accuracy in GUI readiness prediction, compared with 54.13\% for the backbone VLM Qwen3-VL-2B-Instruct~\cite{qwen2025qwen3vl}.
Moreover, we evaluate \tool using industrial GUI testing recordings from Meituan, a major Chinese local-services platform with over 800 million transacting users~\cite{meituan2025q3results}.
In our industrial replay evaluation, \tool reduces mean post-Ready waiting by 94.98\% compared with the existing fixed waiting strategy.
We further analyze representative failure cases to understand the remaining challenges and identify directions for future research.

We summarize the contributions of this paper as follows.
\begin{itemize}
    \item We share industrial observations that identify fixed post-action waiting as a primary source of overhead in GUI testing agents. 
    Our motivation study reveals that over 60\% of the fixed waiting time is wasted after the interface is already ready.

    \item We introduce \tool, an VLM-based adaptive waiting approach that learns when to end waiting from temporal GUI observations.
    We make this approach practical by fine-tuning a lightweight open-source VLM on annotated GUI loading sequences.

    \item Comprehensive experiments show that \tool outperforms existing baselines in detecting GUI loading completion with low runtime overhead. 
    Moreover, our evaluation on Meituan's industrial recordings demonstrates its practical potential for reducing unnecessary waiting, while failure case analysis provides insights for future research.
\end{itemize}

\section{Background}

\subsection{Industrial Practice}

Industrial mobile GUI testing has traditionally relied on automated exploration~\cite{su2017stoat,li2017droidbot,lv2022fastbot2} and manually maintained test scripts~\cite{hu2018appflow,guo2019sara}.
More recently, LLM-based GUI testing agents have been introduced to execute test tasks described in natural language~\cite{hu2024auitestagent,kong2025prophetagent}.
In industrial regression testing, these agents may run across diverse devices and network conditions to cover different user environments~\cite{istqb2019mobile} (\eg AWS Device Farm~\cite{aws2026devicefarm} and Firebase Test Lab~\cite{google2026firebasetestlab}).
Differences in hardware capabilities and network conditions can lead to substantial variations in GUI loading time, ranging from less than a second to tens of seconds.
Unlike random exploration tools, GUI testing agents rely on the outcome of each interaction to plan subsequent actions.
Observing the screen too early risks capturing blank screens, loading spinners, or other intermediate loading states that mislead the agent into incorrect actions.
To avoid such errors, existing GUI agent implementations often adopt configurable fixed waiting times~\cite{wang2024mobileagent,zhang2025appagent,rawles2025androidworld,zhang2025ufo2,zhang2025litewebagent}.

At Meituan, we deploy GUI testing agents~\cite{hu2024auitestagent,wang2025testweaver} to verify mobile services including food delivery, hotel booking, and retail.
Our testing pipeline executes tasks across a shared pool of physical devices, including resource-constrained smartphones.
To accommodate slow-loading interfaces, the configuration examined in this study adopts a fixed 20-second waiting interval after each interaction before the agent captures its next screenshot.
To investigate the overhead introduced by this waiting strategy, we analyze real-world GUI agent executions from Meituan in the following subsection.

\subsection{Motivation Study}
\label{ssec:motivation_study}

To quantify the waiting overhead in industrial GUI testing agents, we investigate the following question:
\textit{How much execution time do GUI testing agents spend on fixed waiting, and how much of that waiting is unnecessary?}

We collect 100 screen recordings of GUI testing agent executions from Meituan's regression pipeline, covering four low-end devices under slow network conditions.
The pipeline adopts a fixed waiting interval of $W=20$ seconds after each interaction.
Following prior work~\cite{feng2023adat}, we focus on click interactions, yielding 468 clicks.
We regard an interface as \textit{Ready} when the primary response to the preceding action has appeared and provides sufficient information for the agent's next observation.
For each click, two annotators independently identify the earliest time when the interface reaches the Ready state, with disagreements resolved by a third reviewer.

We measure two complementary metrics: \textit{Ready latency} and \textit{post-Ready waiting}.
Specifically, we break down the time within the sampled interaction intervals into fixed waiting, model calls, and other time, including action execution.
Ready latency measures the elapsed time from action completion to the moment the interface reaches Ready.
It reflects how quickly the interface becomes sufficient for the agent's next observation.
Post-Ready waiting measures the time between reaching Ready and the end of the 20-second fixed waiting interval.
It is zero when Ready latency is at least 20~seconds.
A higher value indicates more redundant waiting after the interface is sufficient for the agent's next observation.

\begin{figure}[t]
    \centering
    \includegraphics[width=0.9\columnwidth]{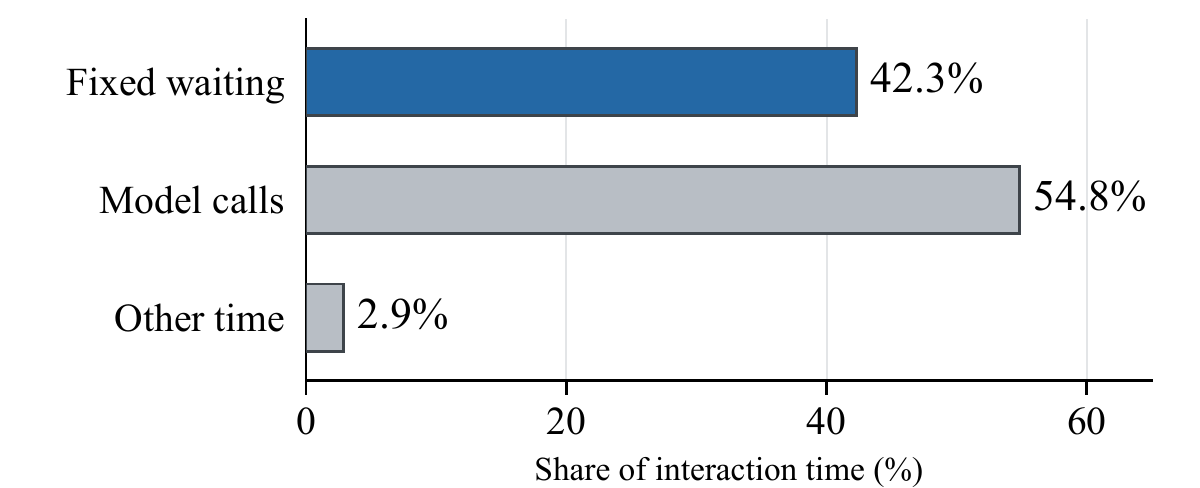}
    \caption{Breakdown of time within the sampled click interaction intervals.}
    \label{fig:runtime_breakdown}
\end{figure}
\begin{figure}[t]
    \centering
    \includegraphics[width=\columnwidth]{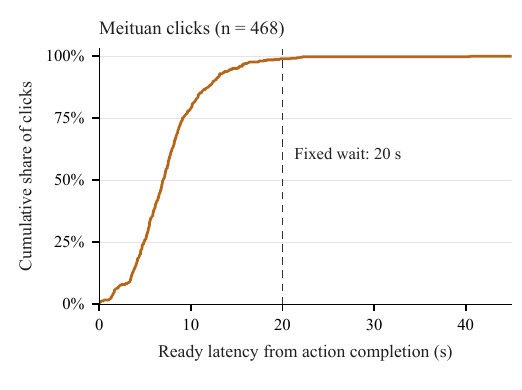}
    \caption{Empirical cumulative distribution of Ready latency for 468 Meituan clicks. The dashed line marks the configured 20-second waiting interval.}
    \label{fig:meituan_ready_latency}
\end{figure}

Figure~\ref{fig:runtime_breakdown} presents the runtime breakdown.
Fixed waiting accounts for 42.3\% of the measured execution time, comparable to the 54.8\% spent on model calls, while other time accounts for 2.9\%.
This indicates that waiting remains a substantial source of execution overhead in LLM-based GUI testing agents.

Figure~\ref{fig:meituan_ready_latency} shows the distribution of Ready latency.
The median latency is 6.97 seconds, substantially shorter than the configured 20-second wait.
At the same time, response times vary widely across interactions.
Overall, 463 out of 468 clicks (98.9\%) reach Ready within 20 seconds, while the remaining five require longer.
Consequently, post-Ready waiting averages 12.58 seconds per click, accounting for 62.9\% of the total fixed waiting time.
Across all 468 clicks, the configured waits total 156~minutes, of which approximately 98~minutes are spent idling after the interface is already sufficient for the agent's next decision.
This reveals substantial unnecessary waiting in the existing industrial testing pipeline.

These findings highlight the limitations of fixed waiting.
Shorter intervals reduce redundant waiting but increase the risk of premature observations, while longer intervals accommodate slow responses at the cost of additional waiting.
Therefore, adaptive waiting is needed to balance testing effectiveness and efficiency across diverse GUI interactions.

\section{Method}
\label{sec:method}

\subsection{Overview}
\label{sec:method_overview}

\begin{figure*}[t]
    \centering
    \includegraphics[width=0.8\textwidth]{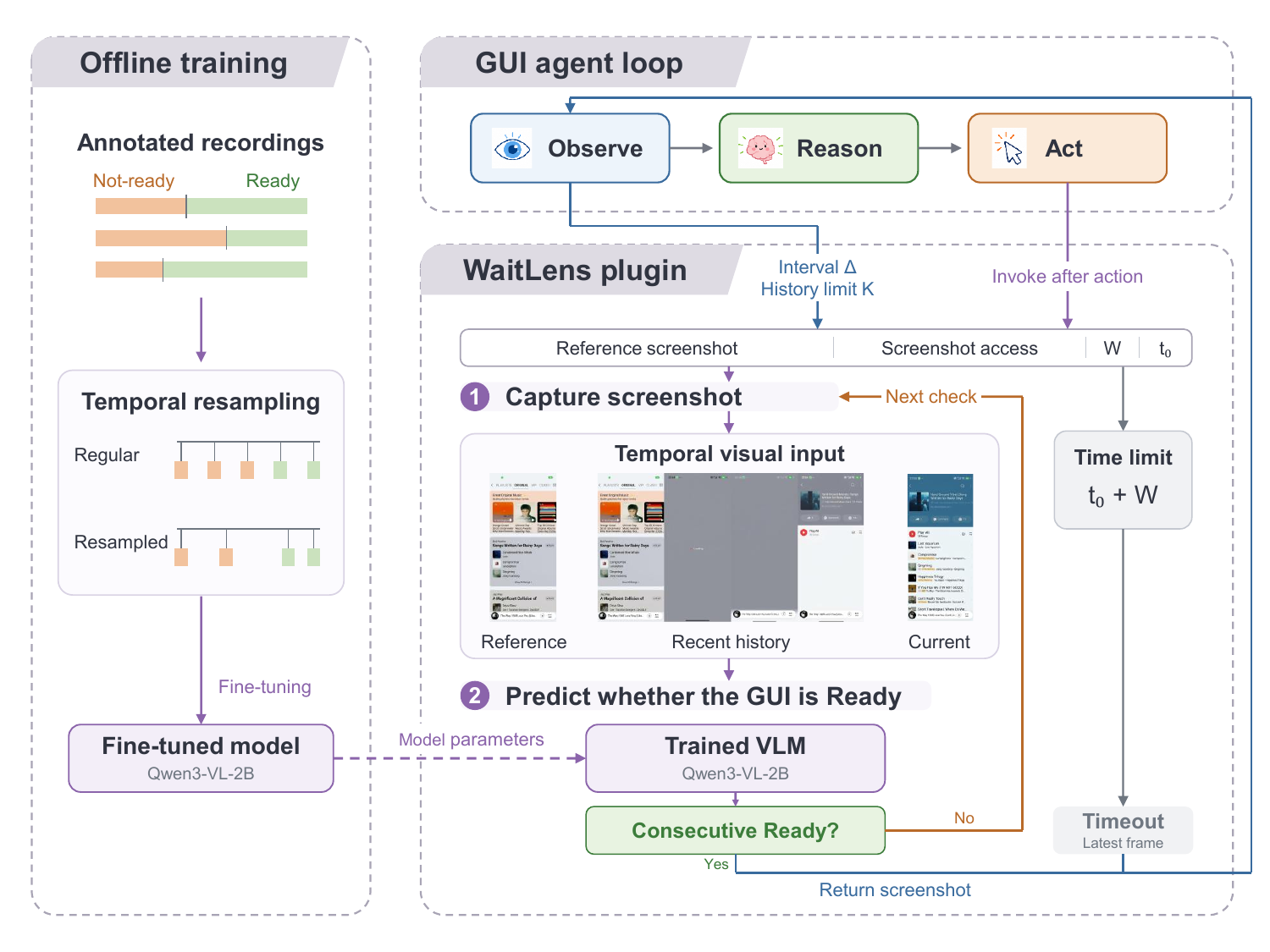}
    \caption{Overview of \tool. Online, the VLM uses bounded temporal visual input to predict whether the GUI has reached the Ready state and control waiting. Regularly sampled and temporally resampled examples are used to train the VLM.}
    \label{fig:method_overview}
\end{figure*}

\tool is a VLM-based adaptive waiting plugin for GUI testing agents.
It replaces the fixed waiting time after each interaction by determining when the interface is ready for the next observation.

As shown in Figure~\ref{fig:method_overview} (right), \tool is invoked immediately after the agent completes an action.
During online waiting, \tool periodically captures the latest screenshot and combines it with the pre-interaction screenshot and a limited number of recent historical screenshots.
The trained VLM uses this temporal visual input to predict whether the GUI has reached the Ready state.
Waiting ends early once the model produces the required number of consecutive \texttt{ready} predictions.
\tool then returns the latest checked screenshot to the agent for its next observation.
Otherwise, it continues checking subsequent screenshots.
If the original waiting budget expires before this condition is met, \tool returns the latest available screenshot.

To train the VLM, we construct labeled visual sequences from annotated interaction recordings and augment them through temporal resampling.
We then fine-tune a lightweight pretrained VLM (Qwen3-VL-2B) using these samples, as illustrated in Figure~\ref{fig:method_overview} (left).

\subsection{Bounded Temporal Visual Representation}
\label{sec:temporal_representation}

\begin{figure*}[t]
    \centering
    \includegraphics[width=\textwidth]{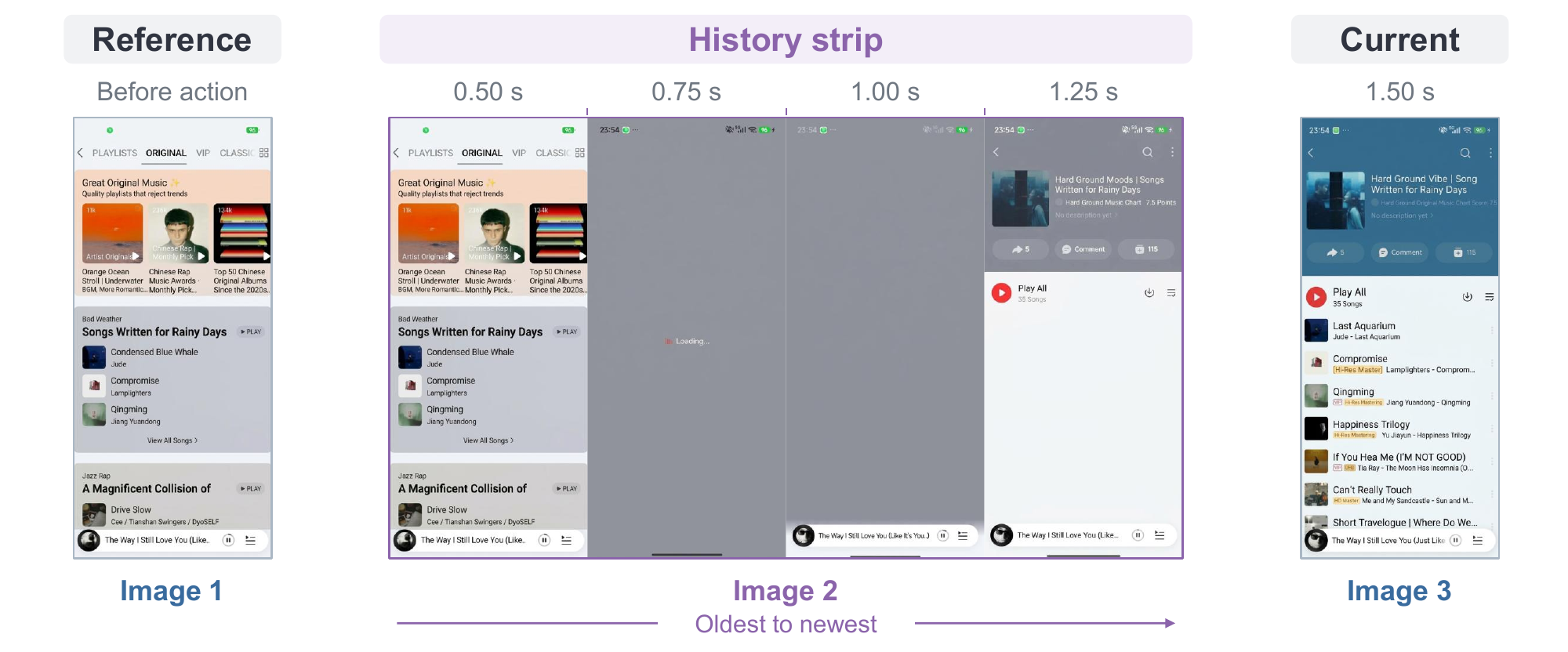}
    \caption{A recorded NetEase Cloud Music interaction illustrating the pre-interaction reference, a chronological strip of four historical screenshots selected by the sliding window, and the current screenshot. Times are relative to action completion; image display heights are normalized for readability.}
    \label{fig:method_visual_input}
\end{figure*}

A single screenshot may not reveal whether the GUI has reached the Ready state, as the GUI evolves over time.
For example, Figure~\ref{fig:method_visual_input} shows that the playlist header appears at 1.25~s, while the song list does not appear until 1.50~s.
Without observing this change, the earlier screenshot may be mistakenly considered Ready.
Therefore, we organize the VLM input using three complementary types of screenshots to capture GUI changes over time.
The pre-interaction screenshot provides a reference for identifying changes triggered by the action.
Historical screenshots capture how the interface evolves, while the current screenshot is used to determine whether waiting should end.

Historical screenshots help the VLM identify recent GUI changes, but retaining every screenshot would increase the input size as waiting continues.
To expose recent interface changes while keeping the input size bounded, we select historical screenshots using a sliding window.
Specifically, let $t_0$ denote action completion, $\Delta$ the screenshot sampling interval, and $K$ the maximum history length.
Screenshots are sampled at regular intervals $t_k=t_0+k\Delta$ ($k\geq1$).
For the current observation at $t_k$, the history retains at most the $K$ most recent preceding screenshots:
\begin{equation}
    H_k=[I(t_{k-N}),\ldots,I(t_{k-1})],
    \qquad N=\min(K,k-1),
    \label{eq:sliding_history}
\end{equation}
where $I(t)$ denotes the screenshot at time $t$.
Thus, the history grows as early screenshots become available ($k \le K$) and slides forward to drop older frames once it reaches the limit of $K$ images.
In the example of Figure~\ref{fig:method_visual_input}, the sampling interval is $\Delta=0.25$~s and the history limit is $K=4$.

After selecting the history, we concatenate its images from oldest to newest into a horizontal strip.
The pre-interaction and current screenshots remain separate, preserving the distinction between contextual evidence and the image being judged.
To preserve interface details while limiting input size, we retain the pre-interaction and current screenshots at higher resolutions and downsample the historical screenshots.
When no history exists ($N=0$), the history strip is omitted.
Finally, we provide these images to the VLM with a prompt specifying their roles and temporal order.
The VLM predicts whether the current interface is \texttt{ready} or \texttt{not\_ready}.

\subsection{VLM Fine-Tuning with Temporal Resampling}
\label{sec:finetuning}

\begin{figure*}[t]
    \centering
    \includegraphics[width=0.8\textwidth]{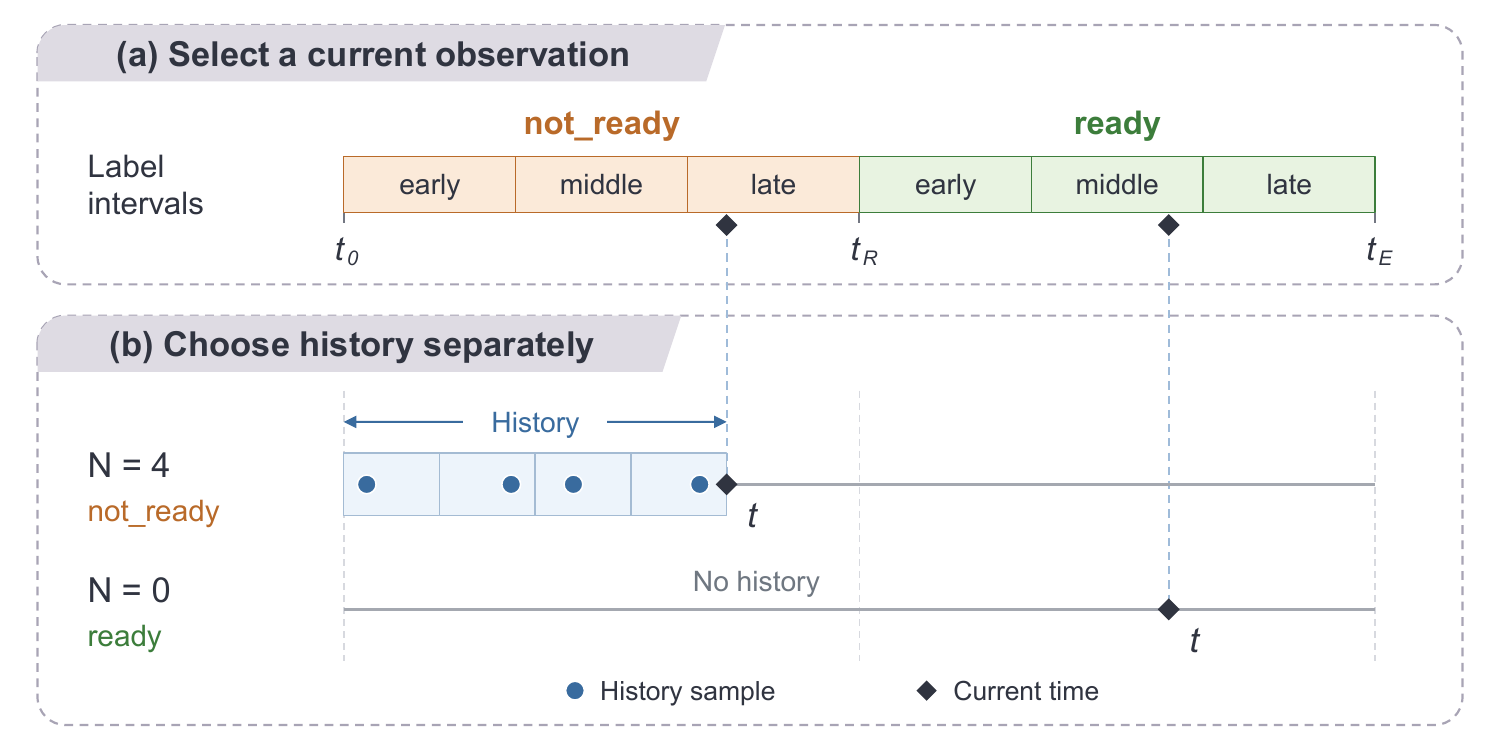}
    \caption{Temporal resampling. (a) Each label interval is divided into three temporal segments from which current observation times are drawn. (b) Given a selected current time $t$, history is sampled from equal-duration strata of $(t_0,t)$, illustrated with $N=4$ and $N=0$.}
    \label{fig:method_sampling}
\end{figure*}

To train the VLM to predict when waiting should end, we construct labeled examples from annotated action recordings $S$.
For each recording, let $t_0$ denote action completion, $t_R$ the annotated earliest Ready time, and $t_E$ the start of the next action or the end of the recording.
Observations before $t_R$ are labeled \texttt{not\_ready}, while those from $t_R$ to $t_E$ are labeled \texttt{ready}.
Following the online procedure described in Section~\ref{sec:temporal_representation}, we sample screenshots at regular intervals of $\Delta$ and pair each current screenshot with the pre-interaction reference and up to $K$ preceding screenshots.
To avoid over-representing slow-loading actions, we retain at most one regularly sampled example per label for each action, forming the baseline set $D_G$.

However, in $D_G$, fixed-interval sampling may introduce a spurious correlation between history length and training labels.
Since history length increases with elapsed time until reaching $K$, early observations tend to have shorter histories than later ones.
As GUIs typically become ready later in the loading process, the VLM may \textit{associate short histories with \texttt{not\_ready} rather than learning from actual GUI changes}.
This correlation may mislead the model when GUI loading speeds vary across interactions and devices.

To break this coupling, we introduce \textit{temporal resampling}, which varies observation times and historical frame spacing while \textit{selecting history length independently of observation time}.
Specifically, we draw on the variability of GUI response speeds across devices and network conditions.
A similar visual progression may unfold quickly on one device and slowly on another, suggesting that temporal compression or stretching can approximate variations in response speed.
For an observer sampling at a fixed interval, these variations change how much visual progress is visible between successive screenshots.
We can construct such temporal variations from a recorded interaction by selecting screenshots at different temporal spacings.

For each action, target label, and history length $N\in\{0,\ldots,K\}$, we generate candidate examples in two steps and collect them in a set $C$.
First, to cover the early, middle, and late stages of each phase, we divide the corresponding label interval into three equal-duration segments.
We cycle through these segments, drawing a current observation time $t$ uniformly within each selected segment, as shown in Figure~\ref{fig:method_sampling}(a).
Second, when $N>0$, we partition the preceding interval $(t_0,t)$ into $N$ equal-duration strata and draw one historical frame uniformly from each stratum, as illustrated in Figure~\ref{fig:method_sampling}(b).
When $N=0$, the history is omitted.
By specifying $N$ independently of $t$, this process synthesizes short or empty histories for \texttt{ready} observations and full histories for \texttt{not\_ready} observations while preserving chronological order.

Finally, we select a subset $C^\star\subseteq C$ to supplement $D_G$ toward balanced coverage across the four groups formed by pairing short ($N<K$) or full ($N=K$) histories with \texttt{ready} or \texttt{not\_ready}.
Within the short-history groups, selection also targets balanced coverage of individual history lengths for each label.
Algorithm~\ref{alg:temporal_resampling} summarizes this procedure.
We then fine-tune the pretrained VLM on the combined training set $D=D_G\cup C^\star$ by minimizing the cross-entropy loss over the target label $y(t)$.

\begin{algorithm}[!tbp]
    \makeatletter
    \renewcommand{\@fs@capt}[2]{{\@fs@cfont #1:} #2\par}
    \makeatother
    \caption{Temporal Resampling}
    \label{alg:temporal_resampling}
    \algrenewcommand{\algorithmicindent}{1em}
    \begin{algorithmic}[1]
    \Require Regularly sampled examples $D_G$; recordings $S$; history limit $K$
    \Ensure Supplemented training set $D$
    \State $C\gets\varnothing$
    \For{each action $s\in S$}
        \State Read $t_0,t_R,t_E$ and the reference from $s$
        \For{each label $y$ and $N\in\{0,\ldots,K\}$}
            \State Split the interval for $y$ into three equal segments
            \State Draw times $T$ by cycling through the segments,
            \Statex \hspace{2em}sampling uniformly within each selected segment
            \For{each selected current time $t\in T$}
                \State $H\gets\varnothing$
                \If{$N>0$}
                    \State Divide $(t_0,t)$ into $N$ ordered strata
                    \State $H\gets$ one uniform draw per stratum
                \EndIf
                \State Form $x$ from the reference, frames at $H$, and $I(t)$
                \State $C\gets C\cup\{(x,y(t))\}$
            \EndFor
        \EndFor
    \EndFor
    \State Select $C^\star\subseteq C$ toward balanced coverage of
    \Statex \hspace{\algorithmicindent}the four history-length and label groups,
    \Statex \hspace{\algorithmicindent}and individual lengths within short-history groups
    \State \Return $D_G\cup C^\star$
    \end{algorithmic}
\end{algorithm}

\subsection{Integration with GUI Agents}
\label{sec:readiness_waiting}

To support integration with diverse agents and applications, we design \tool as a non-intrusive, agent-agnostic plugin.
Specifically, \tool integrates into screenshot-based GUI agents by replacing the fixed waiting period between action execution and the next observation, as shown in the upper of Figure~\ref{fig:method_overview}.
Upon action completion at $t_0$, the agent provides \tool with the pre-interaction screenshot, screenshot access, and its original waiting budget $W$.
The plugin determines when to return the next screenshot without modifying the agent's reasoning or action-execution logic.
Moreover, it operates entirely on screenshots, requiring no application instrumentation, source-code modification, or view-hierarchy access.

During online waiting, \tool samples screenshots at interval $\Delta$ and maintains a history of up to $K$ screenshots.
The waiting budget $W$ sets a deadline at $t_0+W$.
At each check, the fine-tuned VLM predicts whether the current interface is \texttt{ready} using the temporal visual input described in Section~\ref{sec:temporal_representation}.
To reduce prediction fluctuations, \tool supports a configurable threshold $n$ that requires $n$ consecutive \texttt{ready} predictions before returning.
Once this condition is met, \tool ends waiting and returns the latest checked screenshot to the agent.
Otherwise, it continues checking until the original waiting budget $W$ expires.
If no \texttt{ready} prediction is obtained by then, \tool returns the latest available screenshot.
This allows the agent to resume earlier when the GUI is ready while retaining the original waiting budget as a fallback.

\section{Evaluation}
\label{sec:evaluation}

We evaluate \tool using GUI interaction recordings collected from Meituan's industrial testing pipeline.
Our evaluation examines GUI readiness prediction and waiting-policy performance through offline replay.
Specifically, we address the following research questions (RQs):

\begin{enumerate}
    \item \textbf{RQ1:} How accurately and efficiently does \tool predict GUI readiness, and how does temporal resampling affect its performance?
    \item \textbf{RQ2:} How effectively can \tool reduce extra waiting time while minimizing premature returns in industrial GUI testing?
\end{enumerate}

\subsection{Experimental Setup}
\label{sec:evaluation_setup}

\subsubsection{Datasets}
For \textbf{RQ1}, we use 102 annotated interaction recordings covering 12 testing tasks from Meituan.
We split the tasks into training, validation, and test sets with 7, 2, and 3 tasks, respectively.
All recordings and derived observations from the same task are assigned to the same set to avoid data leakage.
Following Section~\ref{sec:finetuning}, we construct 994 training, 700 validation, and 848 test observations using regular sampling and temporal resampling.
To evaluate readiness prediction under both regular sampling and varying observation intervals, the test set contains 424 regularly sampled and 424 temporally resampled observations.

For \textbf{RQ2}, we use a larger dataset collected from Meituan's daily regression testing pipeline to evaluate waiting policies through offline replay.
It contains 1,101 recordings and $M=3{,}868$ post-action GUI response sequences, each serving as one evaluation unit.
Of these, 512 recordings (1,670 interactions) are captured directly from device screens, while 589 recordings (2,198 interactions) are captured by external cameras using physical test rigs.
We geometrically correct the external recordings to align their frames with the screen viewport.
Figure~\ref{fig:case_study} shows examples of the two recording types.
The earliest Ready time for each interaction was annotated by Meituan testing personnel, providing ground truth for evaluating premature returns and post-Ready waiting.
The RQ1 and RQ2 datasets contain no overlapping recordings.

\subsubsection{Baselines and Variants}

For \textbf{RQ1}, we compare \tool with four baselines: pretrained Qwen3-VL-2B-Instruct, Qwen3.8-Max~\cite{qwen2026qwen38max}, AdaT~\cite{feng2023adat}, and a variant of \tool trained without temporal resampling.
The Qwen3-VL-2B-Instruct evaluates the effect of fine-tuning, while Qwen3.8-Max serves as a larger general-purpose VLM baseline.
Both VLMs receive the same temporal visual input as \tool.
For AdaT, we use its released MobileNet classifier with the current screenshot, mapping its fully and partially rendered predictions to \texttt{ready} and \texttt{not\_ready}, respectively.
We also evaluate an ablation variant of \tool. It is trained without temporal resampling, using only $D_G$, whereas the complete model is trained on $D_G\cup C^\star$.

For \textbf{RQ2}, we compare \tool with AdaT and fixed waiting intervals of $w\in\{1,2,3,5,10,20\}$~s.
The 20~second duration is the default configuration in Meituan's testing pipeline.
Both \tool and AdaT use the same maximum waiting budget of $W=20$~s and follow the return and deadline rules in Section~\ref{sec:readiness_waiting}.
We set $n=2$ to reduce prediction fluctuations and the risk of premature returns.

\begin{table*}[!t]
    \centering
    \begingroup
        
    \centering
    \caption{Ready-state classification and model-call latency on the Meituan test set of 848 observations. Boldface highlights the best accuracy, FPR, and latency results.}
    \label{tab:rq1_results}
    \small
    \setlength{\tabcolsep}{9pt}
    \renewcommand{\arraystretch}{1.16}
    \begin{tabular}{@{}cccccc@{}}
        \toprule
        \multirow[c]{2}{*}[-0.4ex]{Model or Variant}
        & \multicolumn{3}{c}{Ready-State Classification (\%)}
        & \multicolumn{2}{c}{Model-Call Latency (s)} \\
        \cmidrule(lr){2-4}\cmidrule(lr){5-6}
        & Accuracy $\uparrow$ & FPR $\downarrow$ & FNR $\downarrow$
        & Median $\downarrow$ & P95 $\downarrow$ \\
        \midrule
        AdaT
        & 49.17 & 100.00 & 0.00
        & \textbf{0.026} & \textbf{0.056} \\
        Qwen3-VL-2B
        & 54.13 & 75.41 & 15.35
        & 0.153 & 0.286 \\
        Qwen3.8-Max
        & 59.55 & 74.94 & 4.80
        & 1.607 & 2.839 \\
        \midrule
        \tool (w/o Resampling)
        & 83.96 & 29.47 & 2.16
        & 0.232 & 0.734 \\
        \tool
        & \textbf{86.08} & \textbf{24.59} & 2.88
        & 0.243 & 0.786 \\
        \bottomrule
    \end{tabular}

    \endgroup
    \par\vspace{1em}
    \begingroup

    \begin{minipage}{0.92\textwidth}
        \centering
        \caption{Waiting-policy results for 3,868 interactions from Meituan's daily regression pipeline (PRR: premature return rate).}
        \label{tab:rq2_replay}

        \small
        \setlength{\tabcolsep}{4.5pt}
        \renewcommand{\arraystretch}{1.16}

        \begin{tabular*}{\linewidth}{@{\extracolsep{\fill}}cccccc@{}}
            \toprule
            \multirow[c]{2}{*}[-0.4ex]{Waiting Policy}
                & \multirow[c]{2}{*}[-0.4ex]{PRR (\%) $\downarrow$}
                & \multicolumn{2}{c}{Actual Waiting Time (s)}
                & \multicolumn{2}{c}{Post-Ready Waiting (s)} \\
            \cmidrule(lr){3-4}\cmidrule(lr){5-6}
            & & Mean $\downarrow$ & P95 $\downarrow$
                & Mean $\downarrow$ & P95 $\downarrow$ \\
            \midrule
            Fixed 20~s & 0.13
                & 20.000 & 20.000 & 17.957 & 19.384 \\
            Fixed 10~s & 2.25
                & 10.000 & 10.000 & 8.027 & 9.384 \\
            Fixed 5~s & 8.04
                & 5.000 & 5.000 & 3.268 & 4.384 \\
            Fixed 3~s & 13.16
                & 3.000 & 3.000 & 1.472 & 2.384 \\
            Fixed 2~s & 23.60
                & 2.000 & 2.000 & 0.640 & 1.384 \\
            Fixed 1~s & 71.79
                & 1.000 & 1.000 & 0.067 & 0.384 \\
            \midrule
            AdaT & 90.80
                & 1.864 & 20.000 & 1.464 & 18.650 \\
            \midrule
            \textbf{\tool} (all) & 15.64
                & 2.222 & 4.023 & 0.901 & 1.558 \\
            \quad \quad Internal ($n=1{,}670$) & 12.87
                & 2.328 & 3.805 & 0.940 & 1.483 \\
            \quad \quad External ($n=2{,}198$) & 17.74
                & 2.142 & 4.163 & 0.871 & 1.638 \\
            \bottomrule
        \end{tabular*}
    \end{minipage}

    \endgroup
\end{table*}

\subsubsection{Metrics}
For \textbf{RQ1}, we treat predicting whether the GUI is Ready as a binary classification task with \texttt{ready} as the positive class.
We report Accuracy, false positive rate (FPR), and false negative rate (FNR).
A false positive classifies an unready interface as \texttt{ready}, which may cause a premature return.
A false negative classifies a Ready interface as \texttt{not\_ready}, leading to additional checks and longer waiting times.
Since the waiting policy may lead to repeated model calls, we also measure the latency of each model call.
In detail, we report the median and 95th percentile (P95) model-call latency.
Both are measured from submitting the prepared visual input to receiving the prediction, including network latency for API-based models.

For \textbf{RQ2}, we use three metrics to evaluate each waiting policy.
First, the premature return rate (PRR) is the proportion of interactions in which the returned screenshot was captured before the GUI reached Ready.
Second, actual waiting time measures the elapsed time from action completion until the policy returns a screenshot, including checking overhead.
Third, extra waiting after Ready measures the time spent waiting after the GUI reaches Ready and is set to zero for premature returns.
We report PRR and the mean and P95 of both time metrics.

\subsubsection{Implementation and Configuration}
We use Qwen3-VL-2B-Instruct~\cite{qwen2025qwen3vl} as the backbone of \tool, with a default sampling interval of $\Delta=0.25$~s and a maximum history length of $K=4$.
The reference and current screenshots are resized to a longest edge of 768 pixels.
Each historical screenshot is resized to a longest edge of 384 pixels, and the historical screenshots are concatenated horizontally.
All screenshots retain their original aspect ratios.

We fine-tune the model using LoRA with rank 16 and $\alpha=32$, and AdamW with a learning rate of $5\times10^{-5}$.
All VLM-based predictors use the same prompt to output \texttt{ready} or \texttt{not\_ready}.
Local VLM inference runs on a single NVIDIA A100 GPU with bfloat16 precision, a batch size of one, and greedy decoding.
Qwen3.8-Max is accessed through its API with temperature set to zero.

\subsection{RQ1: Ready State Prediction and Overhead}
\label{sec:evaluation_rq1}
As shown in Table~\ref{tab:rq1_results}, \tool achieves the highest accuracy of 86.08\% and the lowest FPR of 24.59\% among the evaluated methods.
Compared with Qwen3-VL-2B and Qwen3.8-Max, \tool improves accuracy by 31.95 and 26.53 percentage points, respectively.
Both general-purpose VLMs exhibit high FPRs (75.41\% and 74.94\%), frequently predicting \texttt{ready} for unready interfaces.
AdaT, using its released checkpoint, predicts \texttt{ready} for all test observations, resulting in an FPR of 100\%.
These results highlight the benefit of task-specific fine-tuning for predicting GUI readiness.

The ablation results further show the contribution of temporal resampling.
Incorporating temporal resampling improves overall accuracy by 2.12 percentage points and reduces FPR from 29.47\% to 24.59\%, with only a minor increase in FNR from 2.16\% to 2.88\%.
This trade-off is important for waiting policies because false positives can cause premature returns, while false negatives lead to additional checks and longer waiting times.
These results suggest that temporal resampling helps reduce premature \texttt{ready} predictions with only a small increase in FNR.

Finally, inference latency is also important because the waiting policy may invoke the model repeatedly.
As shown in Table~\ref{tab:rq1_results}, \tool achieves a median latency of 0.243~s and a P95 of 0.786~s per call.
In comparison, Qwen3.8-Max has a median latency of 1.607~s, approximately 6.6 times that of \tool, and a P95 of 2.839~s when accessed through its API.
\tool achieves higher prediction accuracy and lower model-call latency than Qwen3.8-Max.

\subsection{RQ2: Waiting Policies in Industrial Scenarios}
\label{sec:evaluation_rq2}

Beyond evaluating individual predictions in RQ1, RQ2 examines how effectively \tool controls the post-action waiting process in industrial testing.
In live online testing, network fluctuations introduce unpredictable variations in interface loading times, which can cause the agent's subsequent execution trajectory to diverge and make it difficult to fairly compare different waiting policies.
Therefore, to conduct a strictly controlled comparison, we collect 3,868 interaction response processes from Meituan's daily regression pipeline.
We execute all waiting policies on the exact same production response streams and incorporate the measured duration of every check into the waiting timeline.

As shown in Table~\ref{tab:rq2_replay}, Meituan's default 20~s fixed waiting strategy achieves a near-zero PRR of 0.13\%.
But this conservative strategy incurs an average of 17.957~s of extra waiting after the interface reaches Ready.
In comparison, \tool reduces mean actual waiting from 20.000~s to 2.222~s (P95 of 4.023~s).
It also reduces mean extra waiting by 94.98\%, from 17.957~s to 0.901~s (P95 of 1.558~s).
However, these savings come with a higher PRR of 15.64\%, indicating that premature returns remain a challenge.
Only 2.53\% of interactions reach the 20-second fallback deadline.

\begin{figure*}[t]
    \centering
    \includegraphics[width=0.8\textwidth]{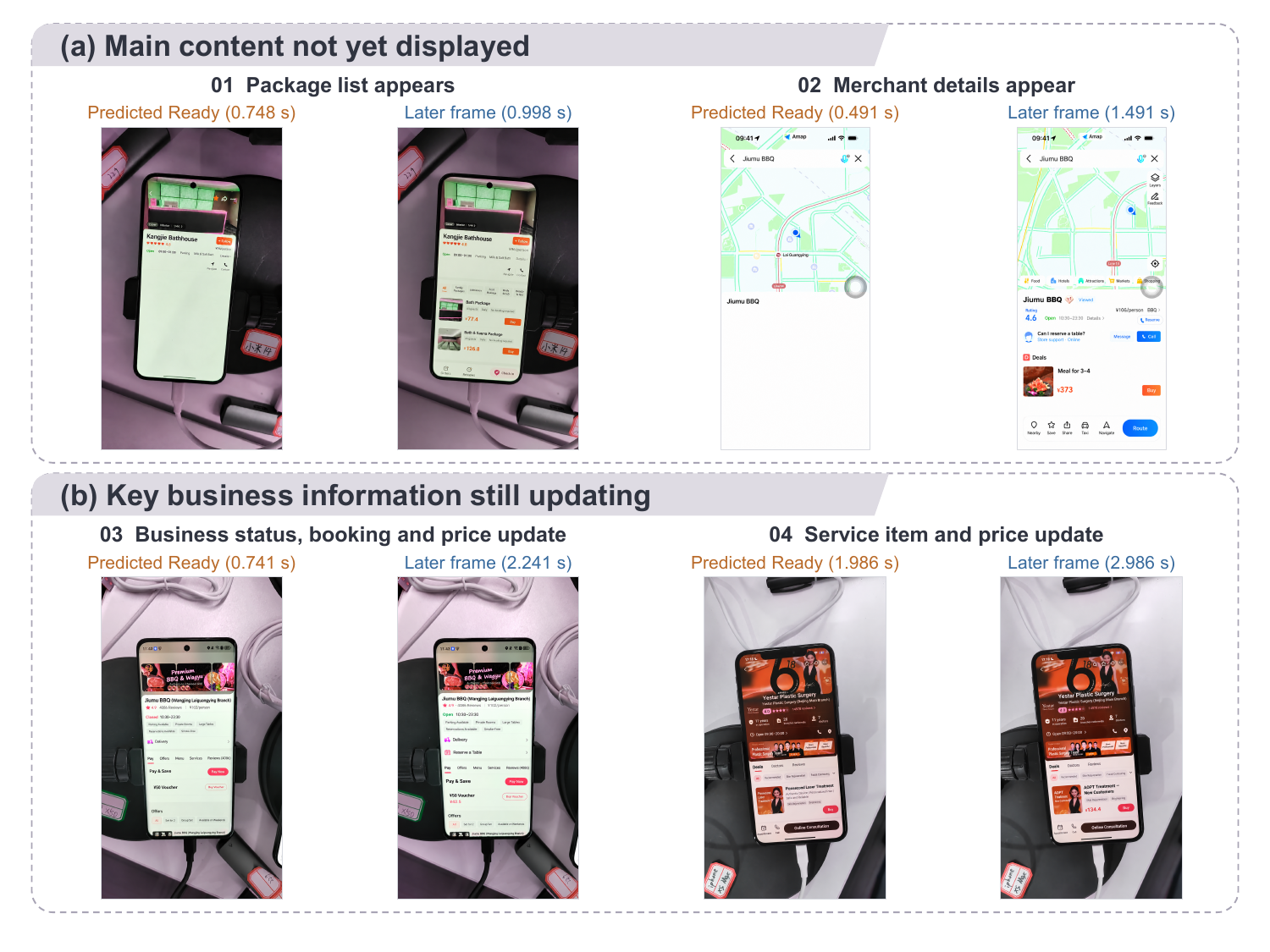}
    \caption{Examples of premature returns. Each pair compares a screenshot judged Ready by \tool with a later screenshot from the same interaction.}
    \label{fig:case_study}
\end{figure*}

Furthermore, comparisons with shorter fixed waiting durations demonstrate that simply reducing a static threshold cannot replicate the efficiency-reliability balance of \tool.
Because interface response times vary widely across interactions, any single static duration faces a rigid trade-off.
Conservative settings of 10~s and 5~s keep PRR relatively low (2.25\% and 8.04\%, respectively).
However, they still incur 8.027~s and 3.268~s of mean extra waiting, respectively, more than three to eight times that of \tool.

Conversely, shortening the fixed wait to 2~s or 1~s to approach \tool's roughly 2~s waiting time causes PRR to surge to 23.60\% and 71.79\%, respectively.
Such aggressive thresholds frequently return unready interfaces to the agent.
Even compared with a 3~s fixed wait (13.16\% PRR), \tool reduces mean actual waiting by 25.9\% and mean extra waiting by 38.8\%, with a slightly higher PRR of 15.64\%.
However, the suitability of a 3~s threshold can only be assessed retrospectively from the observed response times.
In online testing, a fixed threshold selected from historical data may not generalize to changing device and network conditions.
In contrast, \tool dynamically determines when to end waiting based on the observed GUI state, without requiring a predefined threshold tailored to the response time distribution.

Compared with the prior learning-based baseline AdaT, \tool achieves substantially higher reliability in returning ready interfaces.
In industrial testing, an adaptive policy offers little practical value if it reduces waiting time mainly by returning unready interfaces.
AdaT's PRR reaches 90.80\%, meaning that more than nine out of ten interactions return unready interfaces.
By leveraging temporal VLM reasoning, \tool reduces PRR by 75.16 percentage points to 15.64\%.

Finally, we evaluate \tool across internal screen recordings and external camera recordings, achieving PRRs of 12.87\% and 17.74\%, respectively.
Despite not being trained on external camera recordings, \tool maintains comparable performance on this unseen recording modality.
These results demonstrate its potential to generalize across different recording setups in industrial GUI testing.

\subsection{Case Study}
\label{sec:case_study}

In RQ2, \tool returns screenshots before the GUI reaches Ready in 15.64\% of the evaluated interactions.
To better understand these failures, we analyze the premature returns in the Meituan dataset and identify two typical patterns.
Figure~\ref{fig:case_study} compares screenshots judged Ready by \tool with later screenshots from the same interactions.

The first failure pattern involves incomplete content on an otherwise structured page.
Specifically, the page structure is already visible, but the main content is still missing.
In Case 01 of Figure~\ref{fig:case_study}(a), the merchant header is already visible, but the package list and purchase buttons have not appeared.
Similarly, in Case 02, the map and merchant title are displayed, while the merchant details and deals appear later.
Without explicit loading indicators, the missing content may resemble intentionally empty regions, causing the model to predict \texttt{ready} prematurely.

The second pattern involves asynchronous updates to content that initially appears complete.
In Case 03 of Figure~\ref{fig:case_study}(b), the merchant status, booking entry, and voucher price change after \tool predicts \texttt{ready}. 
Similarly, in Case 04, the first service item changes even though the service list and purchase buttons are already visible.
Unlike the first pattern, these pages contain seemingly complete content, making it difficult to determine whether further updates are expected.

Together, these cases illustrate the limitations of relying solely on visual observations to determine GUI readiness.
Without knowing the preceding action and testing goal, \tool may fail to identify missing content or detect updates to important business information. 
A promising direction for future work is to provide the model with the preceding action and testing goal.
This context could help the model identify expected content and relevant updates before deciding whether the GUI is Ready.

\section{Discussion}
\label{sec:discussion}

\subsection{Limitations}
First, \tool relies solely on visual evidence.
Intermediate and completed GUI states may appear similar even with temporal context, leading to premature observations or unnecessary waiting.
The deadline fallback limits waiting when no \texttt{ready} result is available, but the screenshot returned at the deadline may still precede Ready.

Second, the efficiency benefit of adaptive waiting depends on the waiting time it can avoid relative to its checking cost.
Screenshot sampling determines when visual evidence becomes available, while input preparation and model execution introduce further delay.
For fast interactions with short fixed waits, these costs can leave little room for improvement.
Model selection and sampling frequency should therefore account for the host's waiting budget and deployment latency alongside prediction accuracy.

\subsection{Threats to Validity}
Annotators may differ in identifying the earliest ready time for progressively rendered interfaces.
We mitigate this variation through an explicit criterion for the Ready state, two independent annotators, and a third reviewer who resolves disagreements.
Because multiple observations are derived from each recording, all derived observations remain within the source recording's partition.
We further partition Meituan recordings by task group to limit overlap between training and evaluation.

Moreover, our findings may not generalize to all applications, interaction types, and device conditions.
Video replay can compare premature returns and post-Ready waiting across policies using the same recorded response sequence.
Because this sequence remains fixed, replay does not capture how changes in observation timing affect subsequent agent actions.
Establishing effects on task success and overall execution efficiency requires closed-loop agent evaluation.

\section{Related Work}
\subsection{GUI Testing Agents}
Recent advances in LLMs have facilitated the development of autonomous GUI testing agents~\cite{liu2024gptdroid,yoon2024droidagent,ran2024guardian,hu2024auitestagent,kong2025prophetagent,liu2025visiondroid,wang2025testweaver,chen2025memodroid}, reducing the reliance on manually written test scripts~\cite{hu2018appflow,guo2019sara}.
For example, GPTDroid~\cite{liu2024gptdroid} leverages LLMs to guide functionality-aware GUI exploration, while DroidAgent~\cite{yoon2024droidagent} autonomously generates testing goals and interacts with apps to achieve them.
More recently, AUITestAgent~\cite{hu2024auitestagent} enables automated GUI interaction and functional verification based on natural-language test requirements.
Despite their different testing objectives, these agents generally follow an iterative process: observing GUI states, reasoning about subsequent actions, and interacting with apps.
Existing studies primarily focus on improving test generation, exploration, and verification.
In contrast, our work investigates an orthogonal problem: reducing the waiting overhead between consecutive interactions without compromising testing effectiveness.

\subsection{Waiting Strategies for GUI Testing}

Prior studies have investigated various waiting strategies to improve the effectiveness and efficiency of automated GUI testing.
In web GUI testing, WEFix~\cite{liu2024wefix} analyzes browser UI changes to automatically generate explicit wait conditions, while Zhang et al.~\cite{zhang2024robustwaiting} identify expected DOM changes between consecutive actions to determine when tests should proceed.
Moura et al.~\cite{moura2024networkwait} further empirically evaluated different waiting mechanisms in Cypress tests, highlighting the importance of an efficient waiting strategy.

For mobile GUI testing, AdaT~\cite{feng2023adat} employs a CNN-based model to infer GUI rendering states from screenshots and dynamically schedule subsequent test actions.
Its follow-up work, WeReplay~\cite{feng2023wereplay}, applies rendering inference to improve the efficiency of record-and-replay testing in WeChat.
More recently, TurboDroid~\cite{li2026turbodroid} combines static dependency analysis and runtime monitoring to determine when individual GUI widgets are ready for interaction.
In contrast, our work focuses on adaptive waiting for GUI testing agents in industrial practice, where diverse GUI states and heterogeneous devices place greater demands on the generalizability of waiting strategies.

\subsection{VLMs for GUI Understanding}

Recent advances in VLMs have enabled increasingly sophisticated GUI understanding.
Some studies primarily focus on understanding GUI content from static screenshots.
For example, Ferret-UI~\cite{you2024ferretui} enhances the ability of VLMs to recognize, ground, and reason about GUI elements, while Ferret-UI 2~\cite{li2025ferretui2} extends these capabilities across different platforms.
Similarly, SeeClick~\cite{cheng2024seeclick} improves visual GUI grounding to support automated interaction.

More recently, researchers have explored VLMs for understanding dynamic GUI states.
Liang et al.~\cite{liang2026animint} investigate VLMs' ability to interpret UI animations, revealing remaining challenges in understanding their meanings.
PerFrame~\cite{liu2026perframe} employs a VLM to identify loading start and end frames based on recorded screencasts, enabling GUI loading performance monitoring.
In contrast, our work predicts at runtime. It focuses on runtime GUI rendering estimation to guide adaptive waiting for GUI testing agents.

\section{Conclusion}

This paper shows that fixed waiting in GUI testing agents struggles to accommodate substantial variations in interface response times.
Our industrial motivation study reveals that over 60\% of the configured waiting time is spent idling after the interface is already ready for the next observation, resulting in substantial redundant waiting.
To address this, we present \tool, a lightweight VLM-based adaptive waiting approach for GUI testing agents.
By examining temporal GUI changes, \tool predicts whether the GUI is Ready and determines when to end waiting.
Experiments on annotated GUI interaction recordings show that \tool achieves 86.08\% accuracy, compared with 54.13\% for its pretrained backbone, with a median latency of 0.243~s per model call.
Furthermore, our industrial evaluation within Meituan's production testing pipeline demonstrates \tool's real-life usability in substantially reducing redundant waiting while maintaining high observation reliability.

\bibliographystyle{IEEEtran}
\bibliography{references}

\end{document}